\documentclass[twocolumn,pra]{revtex4-1}
\usepackage[latin9]{inputenc}
\usepackage{color}
\usepackage{amsmath}
\usepackage{amssymb}
\usepackage{graphicx}
\usepackage{esint}
\usepackage[unicode=true,pdfusetitle,
 bookmarks=true,bookmarksnumbered=false,bookmarksopen=false,
 breaklinks=false,pdfborder={0 0 0},backref=false,colorlinks=true]
 {hyperref}
\hypersetup{
 colorlinks,linkcolor=red,citecolor=blue}
\usepackage{breakurl}

\makeatletter
\@ifundefined{textcolor}{}
{%
 \definecolor{BLACK}{gray}{0}
 \definecolor{WHITE}{gray}{1}
 \definecolor{RED}{rgb}{1,0,0}
 \definecolor{GREEN}{rgb}{0,1,0}
 \definecolor{BLUE}{rgb}{0,0,1}
 \definecolor{CYAN}{cmyk}{1,0,0,0}
 \definecolor{MAGENTA}{cmyk}{0,1,0,0}
 \definecolor{YELLOW}{cmyk}{0,0,1,0}
}

\usepackage{amsfonts}

\makeatother

\begin{document}

\title{Cavity piezomechanical strong coupling and frequency conversion on
an aluminum nitride chip}

\author{Chang-Ling Zou,$^{1,2}$ Xu Han,$^{1}$ Liang Jiang,$^{2}$ and Hong
X. Tang$^{1,2}$}

\email{hong.tang@yale.edu}

\address{$^{1}$Department of Electrical Engineering, Yale University, New
Haven, Connecticut 06511, USA}

\address{$^{2}$Department of Applied Physics, Yale University, New Haven,
Connecticut 06511, USA}
\begin{abstract}
Schemes to achieve strong coupling between mechanical modes of aluminum
nitride micro-structures and microwave cavity modes due to the piezoelectric
effect are proposed. We show that the strong coupling regime is feasible
for an on-chip aluminum nitride device that is either enclosed by
a three-dimensional microwave cavity or integrated with a superconducting
coplanar resonator. Combining with optomechanics, the piezomechanical
strong coupling permits coherent conversion between microwave and
optical modes with high efficiency. Hence, the piezomechanical system
will be an efficient transducer for applications in hybrid quantum
systems.
\end{abstract}
\maketitle

\section{Introduction}

Efficient conversion between microwave and optical modes has attracted
great attentions recently \cite{Regal2011,Barzanjeh2012,Sillanpaa2014,Tian2015,Kurizki2015}.
The microwave-optical (M-O) interface between superconducting quantum
circuits and optical photons is essential for quantum communications
and distributed quantum computation networks \cite{Kimble2008,Tian2015,Muralidharan2016}.
A high-efficiency M-O converter will enable quantum state transfer
and long-distance commutation between two microwave systems by optical
fibers, without being impacted from microwave thermal noise at room
temperature \cite{Yin2015,Huang2015,Abdi2014}. In addition, the M-O
converter is also useful for single microwave photon level detectors
\cite{Zhang2015a,Barzanjeh2014,Barzanjeh2015}. Generally, the M-O
conversion can be realized by either direct approaches or indirect
approaches. The direct approaches harness the nonlinear optical effects
in dielectrics, such as the electro-optic effect \cite{Tsang2010,Tsang2011,Javerzac-Galy2015}
and the magneto-optic effect \cite{Williamson2014}. The indirect
approaches, on the other hand, exploit intermediate degrees of freedom
to mediate the coupling between microwave and optical photons. For
example, the reversible M-O conversion mediated by phonons has been
proposed and demonstrated experimentally \cite{McGee2013,Andrews2014,Pitanti2015},
where microwave and optical cavities are coupled to a common mechanical
resonator. In addition, recently, the M-O conversion mediated by magnons
has also studied experimentally \cite{Zhang2015,Osada2015}.

The coherent coupling between a microwave cavity and phonon modes
due to piezoelectricity has been demonstrated \cite{OConnell2010,Bochmann2013},
and the piezoelectric effect has also been utilized to actuate mechanical
motion in optomechanical systems \cite{Fan2013,Bochmann2013,Fong2014,Fan2015,Balram2015,Han2015}.
However, integrated systems with both piezoelectric coupling and optomechanical
coupling for M-O conversion have neither been theoretically investigated
nor experimentally demonstrated. In this paper, we propose a strongly
coupled cavity piezo-optomechanical system for the M-O interface,
where mechanical resonators on an aluminum nitride (AlN) chip are
coupled with photonic and microwave cavities simultaneously. To distinguish
from the electromechanics where the capacitive coupling between mechanical
motion and microwave cavities would induce a frequency shift \cite{McGee2013,Andrews2014,Pitanti2015},
we call the linear piezoelectric coupling as piezomechanics. Piezomechanical
systems have several advantages: (i) The frequency of phonon can be
near resonance with the microwave cavity, for example above $10\,\mathrm{GHz}$,
which relaxes the cyrogenic temperature required to cool the system
to the ground state. In contrast, in the flexible electromechanical
system, the phonon frequency is usually orders smaller than that of
microwave photon to enable dispersive microwave photon-phonon coupling
\cite{Andrews2014}, thus requires lower ambient temperature to suppress
thermal phonon excitation. (ii) For near-resonant microwave and mechanical
modes, both the microwave and the optical cavities are within the
resolved-sideband regime, thus the unwanted Stokes process (the parametric
photon-phonon pair generation) is greatly suppressed. (iii) The microwave
and mechanical modes are linearly coupled, hence additional bias DC
field or microwave field is not required. (iv) The systems proposed
here are combined with integrated photonic chips \cite{Han2015},
which are very robust and scalable.

The paper is organized as follows. In Sec. II, we study the basic
features of cavity piezomechanics, develop the Hamiltonian description
of the system, and discuss the possibility of ultrastrong piezomechanical
coupling. Sec. III discusses the implementation in more realistic
physical systems of AlN microstructures coupled to a three-dimensional
(3D) microwave cavity or coupled to a quasi-two-dimensional coplanar
cavity. With practical device parameters, it is estimated that piezomechanical
coupling strength ranging from one to a few hundreds of MHz is feasible.
In Sec. IV, the piezomechanical coupling is combined with the optomechanics,
through which coherent M-O conversion is studied with optimized system
parameters. High conversion efficiency ($\sim90\%$) can be achieved,
mainly limited by the intrinsic optical loss of the material. Sec.
V. summarizes the main results of this paper.

\section{Piezomechanical Coupling}

\begin{figure}
\begin{centering}
\includegraphics[width=1\columnwidth]{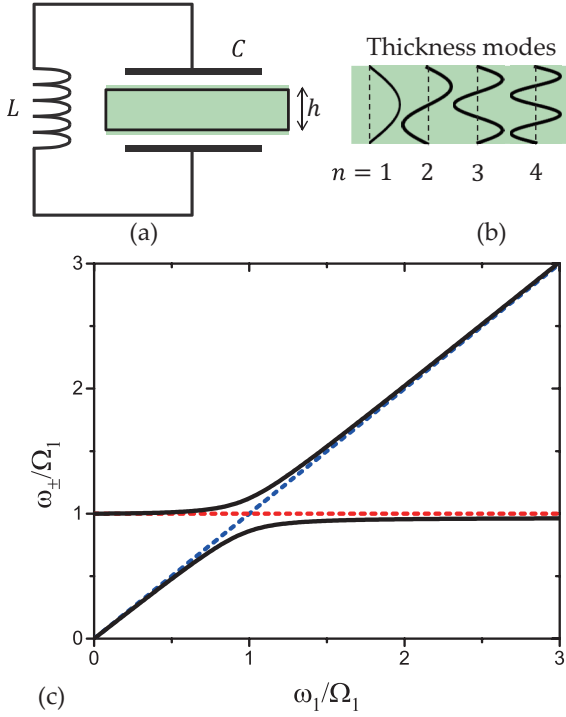}
\par\end{centering}

\protect\caption{(a) Schematic illustration of a piezomechanical system. An LC circuit
is coupled with a thin aluminum nitride film sandwiched in the capacitor.
(b) Distributions of the $z$-direction strain field of thickness
modes. (Only first four orders of modes are shown). (c) Normal mode
frequencies of the piezomechanical system, with the AlN film perfectly
matches the spacing of the capacitor. $\omega_{1}$ and $\Omega_{1}$
are the frequencies of decoupled microwave and fundamental thickness
modes, as indicated by the dashed lines.}

\label{fig-1}
\end{figure}

In general, the internal energy of a piezomechanical system is \cite{yang2005introduction,hashimoto2009rf}
\begin{equation}
U=\frac{1}{2}\int dv(\mathbf{T}\cdot\mathbf{S}+\mathbf{E}\cdot\mathbf{D}),\label{eq:internal energy}
\end{equation}
where $\mathbf{T}$ and $\mathbf{S}$ are the stress and strain fields,
$\mathbf{E}$ and $\mathbf{D}$ are the electric and electric displacement
fields. The piezomechanical interaction can be written in the strain-charge
form as
\begin{eqnarray}
\mathbf{S} & = & \mathbf{s}\cdot\mathbf{T}+\mathbf{d}^{T}\cdot\mathbf{E},\\
\mathbf{D} & = & \mathbf{d}\cdot\mathbf{T}+\mathbf{\epsilon}\cdot\mathbf{E},
\end{eqnarray}
where $\mathbf{s}$, $\mathbf{\epsilon}$, and $\mathbf{d}$ are the
elastic, permittivity, and piezoelectric tensors of the material.
Substitute them into Eq.$\,$(\ref{eq:internal energy}), we obtain
the piezoelectric energy
\begin{eqnarray}
U_{pe} & = & \frac{1}{2}\int dv(\mathbf{T}\cdot\mathbf{d}^{T}\cdot\mathbf{E}+\mathbf{E}\cdot\mathbf{d}\cdot\mathbf{T}).
\end{eqnarray}

The total stress of the mechanical system can be decomposed by the
eigenmodes of the unperturbed system as
\begin{align}
\mathbf{T}(t) & =\frac{1}{\sqrt{2}}\sum_{m}b_{m}\mathbf{T}^{(m)}e^{-i\Omega_{m}t}+h.c.,\\
\mathbf{E}(t) & =\frac{1}{\sqrt{2}}\sum_{n}a_{n}\mathbf{E}^{(n)}e^{-i\omega_{n}t}+h.c..
\end{align}
Here, $b_{m}$ and $a_{n}$ are quantized bosonic operators for the
$m$-th mechanical mode and the $n$-th microwave cavity mode, $\mathbf{T}^{(m)}$
and $\mathbf{E}^{(n)}$ are the corresponding normalized field distributions,
$\Omega_{m}$ and $\omega_{n}$ are the eigenmode frequencies.

The Hamiltonian of the piezomechanical system represented by the Bosonic
operators for phonon $b_{m}^{\dagger}$ and microwave photon $a_{n}^{\dagger}$
is
\begin{eqnarray}
\mathcal{H}_{pm} & = & \sum_{m}\hbar\Omega_{m}b_{m}^{\dagger}b_{m}+\sum_{n}\hbar\omega_{n}a_{n}^{\dagger}a_{n}\nonumber \\
 &  & +\sum_{m,n}\hbar g_{mn}(b_{m}^{\dagger}+b_{m})(a_{n}+a_{n}^{\dagger}),
\end{eqnarray}
with coupling strength
\begin{equation}
g_{mn}=\frac{1}{2}\int dv(\mathbf{T}^{(m)}\cdot\mathbf{d}^{T}\cdot\mathbf{E}^{(n)}+\mathbf{E}^{(n)}\cdot\mathbf{d}\cdot\mathbf{T}^{(m)}).\label{eq-gpm}
\end{equation}
For crystals with hexagon 6mm ($C_{6v}$) symmetry, such as AlN and
GaN \cite{Guy1999}, the piezoelectric coefficient tensor has the
following form
\begin{equation}
\mathbf{d}^{T}=\left(\begin{array}{cccccc}
0 & 0 & 0 & 0 & d_{15} & 0\\
0 & 0 & 0 & d_{24} & 0 & 0\\
d_{31} & d_{32} & d_{33} & 0 & 0 & 0
\end{array}\right),
\end{equation}
where, the first subscript $(1,2,3)$ labels the $(x,y,z)$ components
of the vector field $\mathbf{E}$, while the second subscript $(1,2,...,6)$
labels the $(xx,\thinspace yy,\thinspace zz,\thinspace yz,\thinspace xz,\, xy)$
components of the tensor field $\mathbf{T}$.

Figure$\,$\ref{fig-1}(a) shows a simple cavity piezomechanics system,
where an inductor-capacitor (LC) circuit serves as a microwave resonator,
and a thin aluminum nitride (AlN) film sandwiched by the capacitor
is a film bulk acoustic resonator (FBAR). In the capacitor, the displacement
field $D_{3}$ is constant when the electric fringe effect is neglected.
The value of $D_{3}$ for a single microwave excitation satisfies
$\frac{1}{2}\frac{D_{3}^{2}}{\epsilon_{0}\epsilon_{r}}A_{\mathrm{C}}h+\frac{1}{2}\frac{D_{3}^{2}}{\epsilon_{0}}A_{\mathrm{C}}h_{s}=\frac{1}{2}\hbar\omega_{1}$.
Then we obtain the uniform electric field in the dielectric as
\begin{equation}
E_{3}^{(1)}=\frac{1}{\epsilon_{0}\epsilon_{\mathrm{AlN}}}\sqrt{\frac{\hbar\omega_{1}}{A_{\mathrm{C}}(\frac{h_{s}}{\epsilon_{0}}+\frac{h}{\epsilon_{0}\epsilon_{\mathrm{AlN}}})}}.
\end{equation}
Here, we have only one mode in the LC circuit at $\omega_{1}=1/\sqrt{LC}$,
where $L$ is the inductance and $C=\frac{A_{\mathrm{C}}\epsilon_{0}\epsilon_{\mathrm{AlN}}}{h+h_{s}\epsilon_{\mathrm{AlN}}}$
is the capacitance. $A_{\mathrm{C}}$ is the area of the capacitor,
$h$ and $h_{s}$ are the thickness of AlN and the air spacing in
the capacitor, respectively. $\epsilon_{0}$ is the permittivity of
vacuum, $\epsilon_{\mathrm{AlN}}=8.5$ is the relative permittivity
of the AlN.

The thickness modes of the thin film satisfy the equation \cite{hashimoto2009rf}
$\frac{\partial^{2}u_{3}}{\partial z^{2}}=\frac{1}{v_{l}^{2}}\frac{\partial^{2}u_{3}}{\partial t^{2}}$,
where $u_{3}$ is the displacement in $z$-direction and $v_{l}$
is the longitudinal acoustic wave velocity in the material. The solutions
have the form as $u_{3}^{(n)}(z,t)=[A\sin(k_{n}z)+B\cos(k_{n}z)]e^{-i\Omega_{n}t}$,
with $k_{n}=\frac{\Omega_{n}}{v_{l}}$. Due to the boundary conditions
$T_{3}=0$ at $z=0$ and $z=h$, we have $A\cos(0)-B\sin(0)=0,\, A\cos(k_{n}h)-B\sin(k_{n}h)=0.$
Therefore, $A=0$, $B=1$ and $k_{n}h=n\pi$ for the $n$-th standing
acoustic wave modes as shown in Fig.$\,$\ref{fig-1}(b). Then, the
normalized stress can be solved $T_{3}^{(n)}=\sqrt{2\hbar\Omega_{n}c_{33}/A_{\mathrm{AlN}}h}\sin(k_{n}z)$
by neglecting the effects from the fringe fields, where $c_{33}$
is the elastic constant, $A_{\mathrm{AlN}}$ is the area of the AlN
film. Substituting the expressions of stress and electric fields into
Eq.$\,$(\ref{eq-gpm}), we obtain the coupling strength
\begin{eqnarray}
g_{1n} & = & \xi_{n}\sqrt{\omega_{1}\Omega_{n}},
\end{eqnarray}
where the normalized coupling coefficient
\begin{align}
\xi_{n}= & \frac{1}{\sqrt{2}}\frac{1-\cos n\pi}{n\pi}\frac{A_{\mathrm{ol}}}{\sqrt{A_{\mathrm{C}}A_{\mathrm{AlN}}}}\nonumber \\
 & \times d_{33}\sqrt{\frac{c_{33}}{\epsilon_{0}\epsilon_{r}(\epsilon_{r}h_{s}/h+1)}}.\label{eq:coefficient}
\end{align}
Here, $A_{\mathrm{ol}}<\min\{A_{\mathrm{AlN}},A_{\mathrm{C}}\}$ is
the overlapping area between the AlN film and the capacitor.

The hybridization of the fundamental FBAR mode and the LC mode leads
to new normal modes whose frequencies are
\begin{eqnarray}
\omega_{\pm}^{2} & = & \frac{\omega_{1}^{2}+\Omega_{n}^{2}}{2}\nonumber \\
 &  & \pm\frac{\sqrt{(\omega_{1}^{2}-\Omega_{n}^{2})^{2}+16\xi_{n}^{2}\omega_{1}\Omega_{n}}}{2}.
\end{eqnarray}

From Eq.$\,$(\ref{eq:coefficient}), the expression $(1-\cos n\pi)/n\pi$
indicates that only odd-order mechanical modes can couple with the
LC resonator, and the coupling strength reduces for high order mechanical
modes. The term $A_{\mathrm{ol}}/\sqrt{A_{\mathrm{C}}A_{\mathrm{AlN}}}$
represents the filling factor of the system, which can not exceed
one. Therefore, the largest achievable coupling strength is
\begin{equation}
\xi_{1}=\frac{\sqrt{2}}{\pi}d_{33}\sqrt{\frac{c_{33}}{\epsilon_{0}\epsilon_{\mathrm{AlN}}}},
\end{equation}
for $A_{\mathrm{AlN}}=A_{\mathrm{C}}$ and $h_{s}=0$ that correspond
to unity filling factor. Using the experimentally determined material
constants of AlN \cite{gerlich1986elastic,Guy1999,Dubois1999} $d_{33}=4.0\,\mathrm{pm/V}$
, $\epsilon_{r}=10.4$, $c_{33}=389\,\mathrm{GPa}$, $v_{l}=11\,\mathrm{km/s}$,
$\xi_{1}$ is estimated to be $0.13$. In Fig.$\,$\ref{fig-1}(c),
we plotted the frequencies of normal modes for the ideal setup (solid
lines), comparing with the uncoupled modes (dashed lines). For $\omega_{1}\approx\Omega_{1}$,
the large coupling strength $g_{11}\approx0.13\Omega_{1}$ leads to
the avoided crossing of modes, with a splitting of about $2g_{11}\approx0.26\Omega_{1}$.
This value is even comparable with the resonant frequencies of the
microwave and the mechanical modes, reaching the so-called ``ultrastrong
coupling'' regime \cite{Niemczyk2010,Zhang2014}.

\section{Piezomechanics on AlN chips}

Now we propose possible experimental configurations for realizing
strong piezomechanical coupling. To be compatible with integrated
photonics, we study the piezomechanics on an AlN photonic chip and
using the system parameters from Ref.$\,$\cite{Fong2014,Han2015},
where an AlN thin film ($h=550\,\mathrm{nm}$) is deposited on a silicon
(Si) substrate (thickness is $h_{\mathrm{Si}}=500\,\mathrm{\mu m}$),
with a $2\,\mathrm{\mu m}$ silicon dioxide ($\textrm{SiO}_{2}$)
layer between the AlN and the Si. During the fabrication, AlN microstructures,
such as waveguides and microcavities, can be suspended by etching
away the sacrificial silica layer. The AlN layer, on one hand, confines
the visible or telecom photons due to the high refractive index contrast
to vacuum; on the other hand, it also supports the mechanical thickness
mode to couple to an external 3D microwave cavity or an on-chip microwave
resonator.

In the following calculations, we focus on the fundamental thickness
mode $n=1$ with $\Omega/2\pi=2h/v_{l}=10\,\mathrm{GHz}$ and assume
a mechanical quality factor of $Q_{b}=2\times10^{4}$ at low temperatures
according to experimental results \cite{Fong2014,Han2015}. The corresponding
amplitude decay rate of the mechanical mode is $\kappa_{b}/2\pi\approx0.25\,\mathrm{MHz}$.
For superconducting microwave cavities, an intrinsic quality factor
of $Q_{a}=2\times10^{5}$ is feasible \cite{Megrant2012}, which gives
$\kappa_{a,0}/2\pi\approx0.025\,\mathrm{MHz}$. In the following,
we will demonstrate that the strong coupling that $g_{pm}>\kappa_{b},\kappa_{a,0}$
can be achieved for both 2D and 3D configurations.

\subsection{3D cavity}

\begin{figure}
\begin{centering}
\includegraphics[width=1\columnwidth]{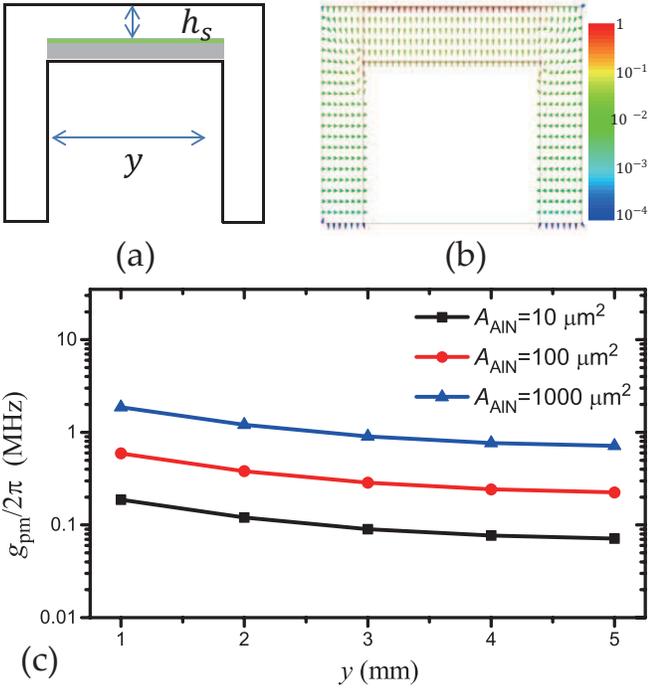}
\par\end{centering}

\protect\caption{(a) Schematic illustration of the 3D cavity piezomechanical system
configuration. (b) The electric field distribution of the cavity for
the square post size $y=3\,\mathrm{mm}$. The chip area matches the
post with $h_{s}=10\,\mathrm{\mu m}$ and $h_{Si}=500\,\mathrm{\mu m}$.
(c) The piezomechanical coupling strength $g_{pm}$ between the fundamental
cavity mode and 1st order thickness mode for different AlN device
areas $A_{\mathrm{AlN}}=10,100,1000\,\mathrm{\mu m}^{3}$.}

\label{fig-3D}
\end{figure}

Figure$\,$\ref{fig-3D}(a) shows the schematic of a system, where
an AlN-on-Si chip is placed in a 3D microwave cavity. For simplicity,
we assume that the AlN microstructure is either waveguide or microring,
with width $w$ and length $l$. To concentrate the electric field
to the photonic chip, we adapt a 3D re-entrant cavity structure \cite{Goryachev2015}
that has a square post in the middle. As shown in Fig.$\,$\ref{fig-3D}(b),
the electric field is greatly enhanced in the small gap between the
post and the top of the cavity.

Based on the numerically simulated electric field distribution, we
calculate the piezomechanical coupling strength $g_{pm}$ using Eq.$\,$(\ref{eq-gpm}).
For simplicity, we assume that the electric distribution is not affected
by the non-uniform AlN layer and only the Si substrate is included
in the numerical model, since the field is mainly determined by the
thick Si substrate. By fixing the size of the Si substrate to match
the size of the square post and introducing $h_{s}=10\,\mathrm{\mu m}$
thick spacing between the chip surface and the cavity wall (Fig.$\,$\ref{fig-3D}(a)),
we calculate $g_{pm}$ with the simulated electric field on the surface
of Si. The results are plotted in Fig.$\,$\ref{fig-3D}(c). Compared
with the case without silicon or air spacing, where $g_{pm}/2\pi=1.3\,\mathrm{GHz}$,
the achievable $g_{pm}$ in 3D cavity is reduced by more than 3 orders,
due to the additional spacing between the capacitor and the reduced filling
factor $F_{3D}=A_{\mathrm{AlN}}/y^{2}$ for practical devices ($g_{pm}\propto\sqrt{F_{3D}}$).
Considering the dielectric constant of Si substrate $\epsilon_{\mathrm{Si}}=12$
and the spacing $h_{s}=10\,\mathrm{\mu m}$, we estimate the best
achievable coupling strength
$g_{pm}/2\pi\approx\frac{\sqrt{2}}{\pi}d_{33}\sqrt{\frac{c_{33}}{\epsilon_{0}\epsilon_{\mathrm{AlN}}}/(\frac{\epsilon_{\mathrm{AlN}}h_{s}}{h}+\frac{\epsilon_{\mathrm{AlN}}h_{\mathrm{Si}}}{\epsilon_{\mathrm{Si}}h}+1)}\Omega\approx85\,\mathrm{MHz}$
for $\Omega/2\pi=10\,\mathrm{GHz}$ with the $A_{\mathrm{AlN}}$ matching
the post size ($F_{3D}=1$). By changing the geometry of the 3D cavity,
we find that the $g_{pm}$ decreases monotonously with increasing
post size, because the cavity mode volume increases with $y$. For
a structure with $w=1\,\mathrm{\mu m}$ and $l=1000\,\mathrm{\mu m}$,
we get $A_{\mathrm{AlN}}=100\,\mathrm{\mu m}^{2}$ and $g_{pm}/2\pi\approx0.59\,\mathrm{MHz}$
for $y=1\,\mathrm{mm}$.

\subsection{Coplanar resonator}

\begin{figure}
\begin{centering}
\includegraphics[width=1\columnwidth]{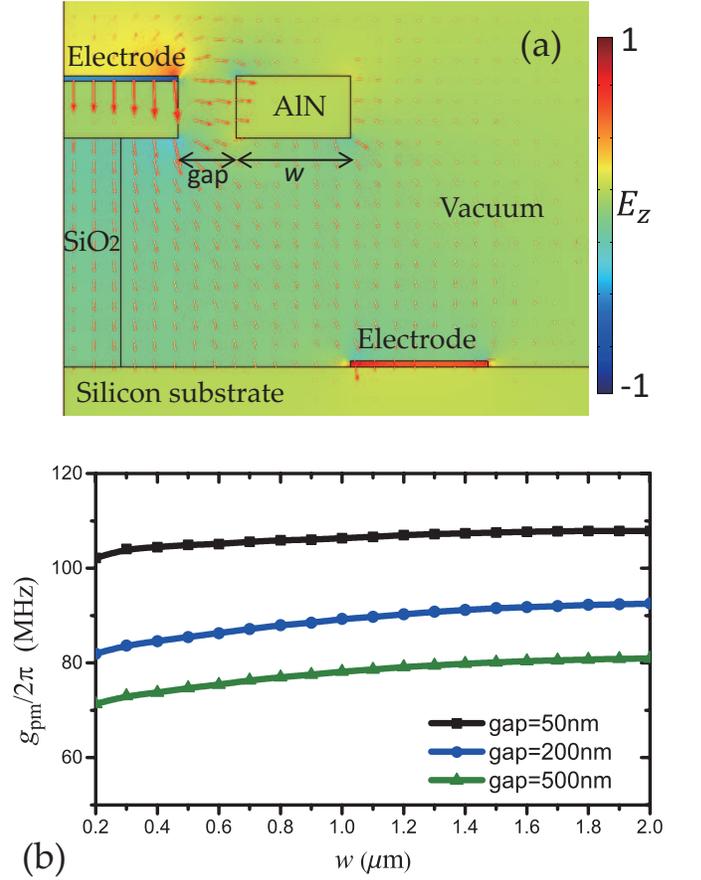}
\par\end{centering}

\protect\caption{(a) The cross section of a suspended AlN waveguide coupled with a
coplanar microwave resonator. The contour plot shows the electric
field $E_{z}$. The arrows indicate the electric field vectors. (b)
The dependence of piezomechanical coupling strength $g_{pm}$ on the
waveguide width $w$ for different AlN to electrode gaps . }

\label{fig-2d}
\end{figure}

Since the 3D microwave cavity has a relatively large mode volume,
we also consider the piezomechanics in a planar geometry with integrated
superconducting resonators. Figure$\,$\ref{fig-2d}(a) shows a schematic
of the proposed structure. To ease the fabrication difficulty, we
avoid the implementation of the parallel capacitor which requires
a buried electrode below the waveguide. Instead, the bottom electrode
is offset from the waveguide to form the ground plane. In addition,
to avoid the metal induced mechanical loss, the top electrode is fabricated
beside the waveguide with a gap between them. This arrangement guarantees
a considerable portion of the out-of-plane electric field in the waveguide,
albeit at a reduced amplitude compared to the parallel plate geometry.

In Fig.$\,$\ref{fig-2d}(a), the electric field distribution for
the coplanar microwave resonator by numerical simulation is plotted.
Although the AlN is not perfectly sandwiched between electrodes, there
is still considerable $E_{z}$-field in the waveguide. With the numerically
solved electric field distribution, the coupling strength $g_{pm}$
is calculated using Eq.$\,$(\ref{eq-gpm}), the results are shown
in Fig.$\,$\ref{fig-2d}(b). Here, we set the microwave cavity length
to be $\lambda_{mw}/4$, where $\lambda_{mw}\approx0.03\,\mathrm{m}$
is the wavelength of microwave. Introducing the filling factor $F_{2D}=4l/\lambda_{mw}$,
we have $g_{pm}\propto\sqrt{F_{2D}}$ when $l<\lambda_{mw}/4$. In
Fig.$\,$\ref{fig-2d}(b), $g_{pm}$ is calculated with $F_{2D}=1$.
$g_{pm}$ reduces with increasing gap, because of the decaying electric
field away from the electrode. When the waveguide width $w$ is increased
from $0.2$ to $2\,\mathrm{\mu m}$, $g_{pm}$ increases due to larger
overlap between the electric field and the AlN structure. As $w$
is further increased, $g_{pm}$ will eventually decrease to zero,
as the size of AlN is much larger than effective volume of microwave
fields. For a structure with $w=1\,\mathrm{\mu m}$ and $l=100\,\mathrm{\mu m}$,
we have $F_{2D}=0.013$. Then, we have $g_{pm}/2\pi\approx12.3\,\mathrm{MHz}$
for a 50-nm gap, which is more than two times larger than the results
obtained in 3D case.

\section{Piezo-optomechanics for frequency conversion}

 Now, we study the strong piezomechanical coupling enhanced M-O
conversion on the AlN chip. Here, we focus on the optomechanics interaction
between the fundamental thickness mechanical mode and the optical
mode in photonic microcavities  \cite{Bochmann2013,Fan2015b}.

\subsection{The optomechanical coupling}

In the optomechanical AlN microstructures, light is confined in the
thick AlN film, and the radiation pressure force on the interface
expands the film in thickness direction. Reversely, the change of
thickness of the film by the mechanical vibration will modify the
boundary condition of electromagnetic field, thus modulate the optical
cavity frequency. Therefore, the optomechanical system can be described
by the Hamiltonian
\begin{equation}
H_{om}=\omega_{c}c^{\dagger}c+\Omega b^{\dagger}b+g_{om}c^{\dagger}c(b^{\dagger}+b).
\end{equation}
Here, $c$ denotes the annihilation operator of optical photon mode.
$g_{om}$ is the vacuum phonon-photon interaction strength and can
be estimated as
\begin{equation}
g_{om}=\frac{\omega_{c}}{n_{eff}}\frac{\partial n_{eff}}{\partial h}2u_{zpf},
\end{equation}
where $u_{zpf}$ is the zero-point fluctuation displacement of the
film, and $n_{eff}\approx1.8$ is the effect index of the waveguide.
Based on numerical simulations, $\frac{\partial n_{eff}}{\partial h}=8.1\times10^{-4}\,\mathrm{nm^{-1}}$,
we obtain
\begin{align}
g_{om}/2\pi & \approx435.6/\sqrt{hwl\times\mathrm{\mu m^{-3}}}\ \mathrm{kHz}.
\end{align}
Here, the optical cavity frequency is $\omega_{c}/2\pi\approx1.95\times10^{14}$
Hz, with the intrinsic material limited loss $\kappa_{c,0}/2\pi=100$
MHz for a quality factor of $Q_{c,0}=10^{6}$.

\subsection{The piezo-optomechanics}

For the purpose of M-O frequency conversion, we need the coherent
conversion between $b$ and $c$ . To compensate the energy difference
between optical photon and microwave phonons, an external laser driving
at frequency $\omega_{d}\approx\omega_{c}-\Omega$ is required. As
$\Omega\gg\kappa_{a,0},\kappa_{b},g_{pm}$, we simply apply the resolved-sideband
approximation that neglects the counter rotating terms, and obtain
the effective Hamiltonian
\begin{equation}
H_{eff}=G_{om}(b^{\dagger}c+bc^{\dagger})+g_{pm}(ba^{\dagger}+b^{\dagger}a),
\end{equation}
with the effective optical photon-phonon coupling strength $G_{om}=\sqrt{N_{d}}g_{om}$,
where $N_{d}$ is the intra-cavity photon number.

The system dynamics incorporating all the interacting modes reads
\begin{align}
\frac{d}{dt}a & =\chi_{a}a-ig_{pm}b-i\sqrt{2\kappa_{a,1}}A_{in}+\sqrt{2\kappa_{a}}\tilde{a},\\
\frac{d}{dt}b & =\chi_{b}b-ig_{pm}a-iG_{om}c+\sqrt{2\kappa_{b}}\tilde{b},\\
\frac{d}{dt}c & =\chi_{c}c-iG_{om}b+\sqrt{2\kappa_{c}}\tilde{c}.
\end{align}
where $\chi_{a}=-i(\omega_{a}-\omega_{mw})-\kappa_{a}$, $\chi_{b}=-i(\Omega-\omega_{mw})-\kappa_{b}$,
$\chi_{c}=-i(\omega_{c}-\omega_{d}-\omega_{mw})-\kappa_{c}$, $\kappa_{a}=\kappa_{a,0}+\kappa_{a,1}$,
$\kappa_{c}=\kappa_{c,0}+\kappa_{c,1}$. Here, $\kappa_{a,0(1)}$
and $\kappa_{c,0(1)}$ denote the intrinsic (external) loss rate of
the microwave and the optical cavity modes. $\tilde{a}$, $\tilde{b}$
and $\tilde{c}$ represents the noise inputs to the system. $\omega_{d}$
is the drive laser frequency and $\omega_{mw}$ is the input microwave
frequency. By neglecting the noises, the steady state conversion efficiency
can be solved as
\begin{align}
T & =\frac{2\kappa_{c,1}c^{\dagger}c}{|A_{in}|^{2}}=\frac{g_{pm}^{2}G_{om}^{2}2\kappa_{a,1}2\kappa_{c,1}}{|G_{om}^{2}\chi_{a}+g_{pm}^{2}\chi_{c}+\chi_{a}\chi_{b}\chi_{c}|^{2}}.
\end{align}
For ideal frequency alignments among input laser and microwave frequencies,
we have $\chi_{a}=-\kappa_{a}$, $\chi_{b}=-\kappa_{b}$, $\chi_{c}=-\kappa_{c}$,
then
\begin{align}
T= &
\frac{\kappa_{a,1}}{\kappa_{a}}\frac{\kappa_{c,1}}{\kappa_{c}}\frac{4\frac{g_{pm}^{2}}{\kappa_{b}\kappa_{a}}\frac{G_{om}^{2}}{\kappa_{b}\kappa_{c}}}{[\frac{G_{om}^{2}}{\kappa_{b}\kappa_{c}}+\frac{g_{pm}^{2}}{\kappa_{b}\kappa_{a}}+1]^{2}}.\label{eq:efficiency}
\end{align}

\subsection{Conversion efficiency}

Although $G_{om}$ can be enhanced by the parametric laser drive,
the ultimate achievable coupling strength is limited by material's
power handling. For instance, as a rule of thumb, the maximum power
delivered to the waveguide without damaging the waveguide is on the
order of $P_{crit}=1\ \mathrm{W/\mu m^{2}}$. In the cavity, the equivalent
circulating power is
\begin{equation}
P=\frac{N_{d}\hbar\omega_{d}/(l/v_{g})}{hw}=\frac{N_{d}\hbar\omega_{d}v_{g}}{hwl},
\end{equation}
where $v_{g}$ is the group velocity of light. The power handling
of the material requires $P<P_{crit}$, corresponding to $N_{d}/hwl\leq5\times10^{4}\ \mathrm{\mu m^{-3}}$,
giving rise to the maximum achievable linear optomechanical coupling
strength
\begin{equation}
G_{om}/2\pi\leq98.9\,\mathrm{MHz}.
\end{equation}
Note that this maximum value is independent of the size of the AlN
photonic structure, regardless of the value of the $g_{pm}$, which
can be adjusted by changing the filling factor of the microstructures.

According to Eq.$\,$(\ref{eq:efficiency}), large $G_{om}$ and $g_{pm}$
are preferred for efficient internal frequency conversion. The optimal
condition requires $\frac{g_{pm}^{2}}{\kappa_{b}\kappa_{a}}\approx\frac{G_{om}^{2}}{\kappa_{b}\kappa_{c}}\gg1$,
which means that $g_{pm}$ should be optimized according to the $G_{om}$,
$\kappa_{a}$ and $\kappa_{c}$. In addition, the other two parameters
-- the extraction ratio $\eta_{a}=\frac{\kappa_{a,1}}{\kappa_{a,0}+\kappa_{a,1}}$
and the input coupling ratio $\eta_{c}=\frac{\kappa_{c,1}}{\kappa_{c,0}+\kappa_{c,1}}$
-- should be optimized for achieving high output efficiency. Here,
the intrinsic loss rates ($\kappa_{a,0}$ and $\kappa_{c,0}$) are
constant while external coupling rates ($\kappa_{a,1}$ and $\kappa_{c,1}$)
are adjustable by varying the structure geometry.

\begin{figure}
\begin{centering}
\includegraphics[width=8cm]{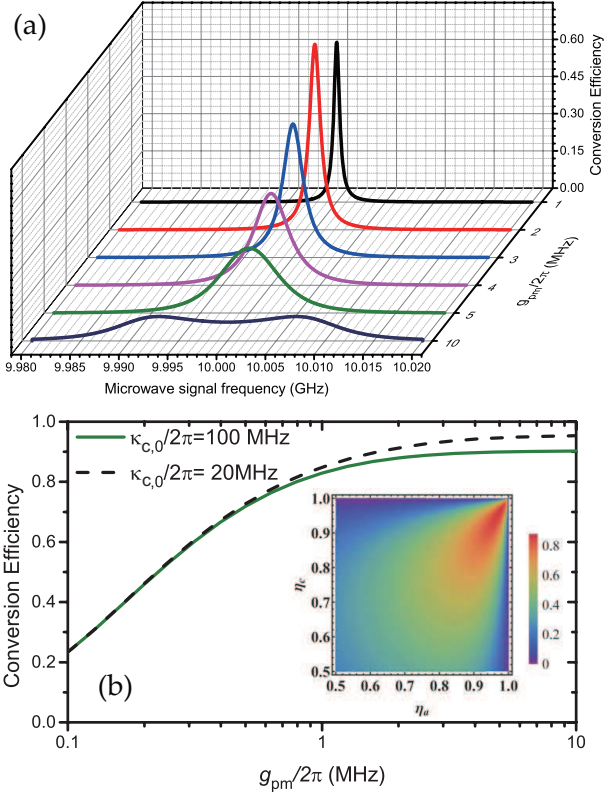}
\par\end{centering}

\protect\caption{(a) The frequency dependence of the conversion efficiency $\xi$ for
different $g_{pm}/2\pi=1,2,3,4,5,10\,\mathrm{MHz}$. (b) The optimal
conversion efficiency for a given $g_{pm}$, by optimizing cavity
extraction ratios $\eta_{a}$ and $\eta_{c}$. The solid (dashed)
lines correspond to intrinsic optical dissipation rates $\kappa_{c,0}/2\pi=100\,(20)\,\mathrm{MHz}$.
The inset shows the conversion efficiency against $\eta_{a}$ and
$\eta_{c}$, with $g_{pm}/2\pi=2\,\mathrm{MHz}$. Other parameters:
$g_{om}/2\pi=98.9\,\mathrm{MHz}$, $\omega_{a}=\omega_{b}$, $\omega_{d}=\omega_{c}-\omega_{b}$,
$\{\kappa_{c,0},\kappa_{b},\kappa_{a,0}\}/2\pi=\{100,0.25,0.025\}\,\mathrm{MHz}$. }

\label{fig-conversion}
\end{figure}

Figure$\,$\ref{fig-conversion}(a) shows the microwave signal conversion
efficiency against the input signal frequency for various piezomechanical
coupling strength $g_{pm}$, with fixed $\eta_{a}=\eta_{c}=0.9$.
At increased $g_{pm}$, the bandwidth of the conversion increases,
while the best efficiency is obtained for optimal $g_{pm}$. From
the spectrum for $g_{pm}/2\pi=10\,\mathrm{MHz}$, we can see two peaks
due to the strong coupling. These results indicate that for different
$g_{pm}$, we should choose optimal $\eta_{a}$ and $\eta_{c}$ for
best conversion efficiency. The inset of Fig.$\,$\ref{fig-conversion}(b)
plots the conversion efficiency as a function of $\eta_{a}$ and $\eta_{c}$
with fixed $g_{pm}/2\pi=2\,\mathrm{MHz}$. It can be seen that the
optimal conversion efficiency $T=0.88$ is realized at $\eta_{a}=0.976,\,\eta_{c}=0.962$.

Therefore, we numerically solve the optimal conversion efficiency
$T$ for on-resonance microwave signal and different $g_{pm}$. The
result is shown as solid curve in Fig.$\,$\ref{fig-conversion}(b),
the conversion efficiency monotonously increases with $g_{pm}$ and
saturated to $T\approx0.9$ when $g_{pm}/2\pi>3\,\mathrm{MHz}$. By
introducing the intrinsic cooperativities $C_{om}=\frac{G_{om}^{2}}{\kappa_{b}\kappa_{c,0}}$
and $C_{pm}=\frac{g_{pm}^{2}}{\kappa_{b}\kappa_{a,0}}$, we can asymptotically
solve the optimal condition for Eq.$\,$(\ref{eq:efficiency}) as
$\kappa_{a,1}\approx\kappa_{a,0}C_{pm}/\sqrt{C_{om}}$ and $\kappa_{c,1}\approx\kappa_{c,0}\sqrt{C_{om}}$
for $C_{pm}\gg C_{om}\gg1$. In this case, the optimal achievable
conversion efficiency is
\begin{align}
T_{\mathrm{sat}} & \approx\frac{C_{om}}{\left(\sqrt{C_{om}+1}+1\right)^{2}}\nonumber \\
 & \approx1-\frac{1}{\sqrt{C_{om}}}+\mathcal{O}\left(\frac{1}{C_{om}}\right),
\end{align}
which is insensitive to the $g_{pm}$. Therefore, the efficiency is
saturated due to the intrinsic optical quality factor limited $C_{om}\sim400$,
leading to $T_{\mathrm{sat}}\approx0.9$. If we increase the intrinsic
optical quality factor by 5 times, i.e. $\kappa_{c,0}/2\pi=20\,\mathrm{MHz}$,
the saturated efficiency can be increased to $T_{\mathrm{sat}}\approx0.95$
which agrees with the numerical results (Dashed line in Fig.$\,$\ref{fig-conversion}(b)).

With reasonable parameters $g_{pm}/2\pi=5\,\mathrm{MHz}$, the optimum
conversion can be obtained with $\eta_{a}=0.995$ and $\eta_{c}=0.954$.
The effective frequency conversion efficiency is $0.9$, with a bandwidth
about $5.4\,\mathrm{MHz}$. If the ambient temperature is $2\,\mathrm{K}$,
the corresponding thermal excitation at $10\,\mathrm{GHz}$ is $n_{th}=3.6$,
and the added noise during the microwave to optical frequency conversion
is $n_{add}\approx0.22$ \cite{Andrews2014}.

\section{Conclusion}

In conclusion, the piezomechanical strong coupling is proposed and
investigated using practical device parameters. The numerical simulations
show that strong coupling can be achieved for microstructures in an
AlN chip coupled to microwave cavity photons in both three-dimensional
microwave cavities and planar superconducting resonators. Leveraging
the piezomechanical strong coupling will lead to greatly enhanced
microwave to optical frequency conversion. With practical parameters,
we show that the optimal conversion efficiency can approach $90\%$,
with a bandwidth exceeding $5\,\mathrm{MHz}$ and added noise below
$0.22$. Compared to other electromechanical schemes, the piezo-optomechanical
system has several advantages and is very promising for experiments.
Thus, piezomechanics in a photonic chip is a promising platform for
building future hybrid quantum technologies and integrated photonics.
\begin{acknowledgments}
We thank L. Fan for discussions. We acknowledge the support from Laboratory
of Physical Sciences (LPS), Air Force Office of Scientific Research
(AFOSR) MURI program, and DARPA ORCHID program. L.J. was also supported
by the ARL-CDQI, ARO, Alfred P. Sloan Foundation, and the Packard
Foundation.
\end{acknowledgments}

\end{document}